\documentclass[ reprint,
superscriptaddress,
 amsmath,amssymb,
 aps,nofootinbib,
 prd
,onecolumn
]{revtex4-2}

\usepackage{graphicx}
\usepackage[usenames,dvipsnames]{xcolor}
\definecolor{bluecite}{HTML}{0875b7}
\usepackage[colorlinks=true,urlcolor=Emerald,linkcolor=black,citecolor=bluecite]{hyperref}
\usepackage{acronym}
\usepackage{MnSymbol}
\usepackage{physics}
\usepackage{setspace}

\newcommand{\eg}{{\textit{e.g.}}}
\newcommand{\ie}{{\textit{i.e.}}}

\def \e{\text{e}}
\def \M{\mathcal{M}}
\def \S{\mathcal{S}}
\def \A{\mathcal{A}}

\begin{document}

\title{Relative entropy of single-mode squeezed states in Quantum Field Theory}

\author{Marcelo S.  Guimaraes}
\email{msguimaraes@uerj.br}
\affiliation{UERJ - Universidade do Estado do Rio de Janeiro,	Instituto de Física - Departamento de Física Teórica - Rua São Francisco Xavier 524, 20550-013, Maracanã, Rio de Janeiro, Brazil}

\author{Itzhak Roditi}
\email{roditi@cbpf.br}
\affiliation{CBPF - Centro Brasileiro de Pesquisas Físicas, Rua Dr. Xavier Sigaud 150, 22290-180, Rio de Janeiro, Brazil}

\author{Silvio P. Sorella}
\email{silvio.sorella@fis.uerj.br}
\affiliation{UERJ - Universidade do Estado do Rio de Janeiro,	Instituto de Física - Departamento de Física Teórica - Rua São Francisco Xavier 524, 20550-013, Maracanã, Rio de Janeiro, Brazil}

\author{Arthur F. Vieira}
\email[]{arthurfvieira@if.ufrj.br}
\affiliation{Universidade Federal do Rio de Janeiro (UFRJ),	Instituto de Física, 21.941-909, Rio de Janeiro, Brazil}
\affiliation{UERJ - Universidade do Estado do Rio de Janeiro,	Instituto de Física - Departamento de Física Teórica - Rua São Francisco Xavier 524, 20550-013, Maracanã, Rio de Janeiro, Brazil}

\date{\today}

\begin{abstract}
Utilizing the Tomita-Takesaki modular theory, we derive a closed-form analytic expression for the Araki-Uhlmann relative entropy between a single-mode squeezed state and the vacuum state in a free relativistic massive scalar Quantum Field Theory within wedge regions of Minkowski spacetime. Similarly to the case of coherent states, this relative entropy is proportional to the smeared Pauli-Jordan distribution. Consequently, the Araki-Uhlmann entropy between a single-mode squeezed state and the vacuum satisfies all expected properties: it remains positive, increases with the size of the Minkowski region under consideration, and decreases as the mass parameter grows. 
\end{abstract}
\maketitle

\tableofcontents

	\section{Introduction}\label{sec:intro}

Quantum information theory has emerged as a powerful framework for analyzing fundamental aspects of quantum mechanics \cite{nielsen2010quantum}, particularly within the context of relativistic Quantum Field Theory (QFT) \cite{Faulkner:2022mlp,Witten:2018zxz,Anastopoulos:2022owu}. A central concept in this framework is the quantum relative entropy, which quantifies the distinguishability between quantum states and plays a crucial role in quantum statistical mechanics, thermodynamics (see, \eg, \cite{Floerchinger:2020ogh,schrofl2024relative}), and the study of entanglement in high-energy physics \cite{Nishioka:2018khk,Witten:2018zxz} (see \cite{Holzhey:1994we,Calabrese:2004eu,Calabrese:2009qy} for early works and \cite{Casini:2022rlv,Berges:2017hne,Schrofl:2023hnz,Katsinis:2023hqn,Hollands:2019czd,DAngelo:2021yat,Ciolli:2021otw,Galanda:2023vjk,Garbarz:2022wxn} for recent accounts).

In this work, we focus on the Araki-Uhlmann relative entropy \cite{araki1975inequalities,Araki:1976zv,Uhlmann:1976me}, a generalization of quantum relative entropy to the setting of von Neumann algebras. This quantity is particularly well-suited for relativistic QFT, where the algebraic approach provides a natural description of quantum states in different spacetime regions \cite{Haag:1992hx,Summers:2003tf}. Specifically, we study the relative entropy for single-mode squeezed vacuum states, a class of states that play a pivotal role in quantum optics \cite{Agarwal}, Bose-Einstein condensates \cite{Dalfovo:1999zz} and QFT. Single-mode
squeezed states are particularly significant in phenomena such as particle production in curved spacetime \cite{parker1969quantized,birrell1984quantum}, the Unruh effect \cite{Ahn:2006jw,Pan:2023tqb,Scully:2022pov}, relativistic Quantum Information \cite{MohammadiMozaffar:2024uiy}, Quantum Chromodynamics \cite{Blaschke:1996dp} as well as in the study of the Bell-CHSH inequality \cite{Summers:1987fn,Summers:1987squ,Guimaraes:2025xij,DeFabritiis:2023llu}.

A key property of single-mode squeezed states is that they can be obtained from the vacuum state via a unitary transformation defined by a single-mode squeezing operator. This feature, when combined with the modular theory of Tomita-Takesaki \cite{Takesaki:1970aki,Witten:2018zxz} and the Bisognano-Wichmann results \cite{Bisognano:1975ih,bisognano1976duality}, allows us to evaluate the Araki-Uhlmann relative entropy explicitly. Indeed, the Tomita-Takesaki modular theory provides a rigorous mathematical framework for understanding modular automorphisms in QFT, while the Bisognano-Wichmann theorem characterizes the modular structure of wedge-localized quantum field theories in Minkowski spacetime. By leveraging these tools, we derive a closed-form analytical expression for the relative entropy of a single-mode squeezed vacuum state in a free relativistic massive scalar QFT and analyze its behavior in wedge regions of Minkowski spacetime (see \cite{Katsinis:2023hqn,Katsinis:2024sek} for recent numerical investigations of squeezed states in QFT).

Our main result is that the Araki-Uhlmann relative entropy between a single-mode squeezed state and the vacuum state is proportional to the smeared Pauli-Jordan distribution. Consequently, the relative entropy exhibits the same fundamental properties observed in the case of coherent states \cite{Witten:2018zxz,Ciolli:2019mjo,Casini:2019qst,Frob:2024ijk,Guimaraes:2025cqt}: it remains positive, increases with the size of the Minkowski region under consideration, and decreases as the mass parameter grows. These findings provide deeper insight into the structure of entanglement and information-theoretic aspects of single-mode squeezed states in QFT, reinforcing the role of relative entropy as a key quantity in the study of quantum correlations in relativistic settings.

The paper is structured as follows: Sec. \ref{sec.background} provides the general properties of single-mode squeezed states for a relativistic scalar QFT. In Sec. \ref{Autt}, we review the necessary background on the modular theory of Tomita-Takesaki as applied to the Araki-Uhlmann relative entropy. In Sec. \ref{comp}, we present the computation of the relative entropy between a single-mode squeezed state localized in the right wedge and the vacuum state. Sec. \ref{conc} collects our conclusions. Finally, Appendices \ref{appA} and \ref{appB} summarize key notions of the canonical quantization of the massive scalar field and provide additional technical details about single-mode squeezed states.

\section{Properties of the single-mode squeezed states in relativistic scalar QFT}\label{sec.background}

To begin, we specify the von Neumann algebra ${\cal M}$ that will be used throughout the discussion. For this purpose, we construct the algebra using the Weyl operators
\cite{Summers:1987fn,Summers:1987squ,DeFabritiis:2023tkh}
\begin{equation} 
{\cal M} = \{\; W_g = e^{i \varphi(g)} \,, \; supp(g) \subset {\cal W}_R \;\}{"} \;, \label{vnM}
\end{equation}
where $g(t,x)$ is a smooth test function with support contained in the right wedge ${\cal W}_R =\{(t,x)\,, x\ge |t| \}$. The symbol ${"}$ in expression \eqref{vnM} means the bicommutant and $\varphi(g)$ stands for the smeared scalar quantum field (see Appendix~\eqref{appA}), 
\begin{eqnarray}
\varphi(t,x) & = & \int \; \frac{d k}{2 \pi} \frac{1}{2 \omega_k} \left( e^{-ik_\mu x^\mu} a_k + e^{ik_\mu x^\mu} a^{\dagger}_k \right) \;, \nonumber \\
\varphi(g) &=& \int d^2x \; g(t,x) \varphi(t,x) = a_g + a^{\dagger}_g \;, \nonumber \\
a_g & = &  \int \; \frac{d k}{2 \pi} \frac{1}{2 \omega_k} g^{*}(\omega_k,k) a_k \;, \qquad a_g^{\dagger}  =  \int \; \frac{d k}{2 \pi} \frac{1}{2 \omega_k} g(\omega_k,k) a^{\dagger}_k  \;, \label{cqf}
\end{eqnarray} 
where $\omega_k  = k^0 = \sqrt{k^2 + m^2}$ and $g(\omega_k,k)$ is the Fourier transform of $g(t,x)$. The smeared creation and annihilation operators $(a_g, a^\dagger_g)$ obey the following relation 
\begin{equation} 
[a_g, a^\dagger_{g'}] = \langle g | g'\rangle = \int \; \frac{d k}{2 \pi} \frac{1}{2 \omega_k} g^{*}(\omega_k,k) g'(\omega_k, k) \;, \label{isp}
\end{equation}
that is, $\langle g | g'\rangle$ is the Lorentz-invariant inner product, see Appendix~\eqref{appA}. When rewritten in configuration space, $\langle g | g'\rangle$ takes the form 
\begin{equation} 
\langle g | g'\rangle =\frac{i}{2} \Delta_{PJ}(g,g') + H(g,g') \;, \label{cfs}
\end{equation}
where $\Delta_{PJ}(g,g') $ and $H(g,g')$ stand, respectively, for the Pauli-Jordan and Hadamard smeared distributions, App.~ \eqref{appA}.

Let us introduce the vacuum state $|\Omega\rangle$ defined by 
\begin{equation} 
a_g |\Omega \rangle  = 0 \;, \forall g {\;}\qquad supp(g) \subset {\cal W}_R \;. \label{vc} 
\end{equation}
From the Reeh-Schlieder theorem\footnote{See \cite{Haag:1992hx}.}, one learns that $ |\Omega \rangle$
 is cyclic and separating for the von Neumann algebra ${\cal M}$, Eq.~\eqref{vnM}. We are now ready to introduce the squeezed state $|\Psi \rangle$ for a scalar QFT, namely 
 \begin{equation} 
 |\Psi\rangle = {\cal S}_f |\Omega\rangle \;, \label{sqs}
 \end{equation}
where $ {\cal S}_f $ is the unitary operator\footnote{The operator defined in Eq.~(\ref{sqop}) represents a single-mode squeezing operator. In a more general context, $\mathcal{A}$ denotes a two-mode operator, which can be formulated by incorporating a two-mode kernel $f(p,q)$. This operator is expressed as
\begin{eqnarray}
    \S_f&=&\e^{i\A_f},\nonumber\\
    \A_f&=&\int\frac{dp}{2\pi} \frac{dq}{2 \pi} \frac{1}{2\omega_p}\frac{1}{2\omega_q}f(p,q)\,a^\dagger_p a^\dagger_q+\textrm{h.c.},\label{twomode}
\end{eqnarray}
where h.c. denotes the Hermitian conjugate. The algebraic structure arising from Eq.~(\ref{twomode}) is more intricate than that of Eq.~(\ref{sqop}); however, the computational approach follows a similar methodology.}
\begin{equation}
 {\cal S}_f  = e^{i A_f} \;, \qquad A_f = a^2_f + (a^\dagger_f)^2 \;, \qquad   {\cal S}_f  {\cal S}_f^\dagger = {\cal S}_f^\dagger  {\cal S}_f =1 \;. \label{sqop}
\end{equation}
As in quantum mechanics \cite{Agarwal}, the three operators $(a_f^2, {a_f^\dagger}^2, a_f^\dagger a_f)$ form a closed algebra. More precisely, 
\begin{eqnarray} 
[a^\dagger_f a_f, a^2_f] & =&  - 2 ||f||^2 a^2_f \;, \nonumber \\
\left[ a^\dagger_f  a_f, {a^\dagger_f}^2 \right] & = & 2 ||f||^2 {a^\dagger_f}^2 \;, \nonumber \\
\left[ a^2_f, {a^\dagger_f}^2 \right] & = & 4 ||f||^2 a^\dagger_f a_f + 2 ||f||^4 \;, \label{clalg}
\end{eqnarray}
where $||f||^2$ is the norm induced by the Lorentz-invariant inner product, {\it i.e.}, $||f||^2 = \langle f|f\rangle$. Relation \eqref{clalg} enables us to rewrite the operator ${\cal S}_f$ in the useful form 
\begin{equation} 
e^{i( a^2_f + {a^\dagger_f}^2) } = {\cal C}_f \; e^{i \alpha_f {a^\dagger_f}^2} \; e^{i \beta_f a^\dagger_f  a_f} \; e^{i \gamma_f a^2_f} \;, \label{des}
\end{equation}
for appropriate values of the parameters $({\cal C}_f, \alpha_f, \beta_f, \gamma_f)$. To derive Eq.~\eqref{des}, we follow an approach similar to that in quantum mechanics by introducing an interpolating parameter $s$, {\it i.e.} 
\begin{equation} 
e^{is( a^2_f + {a^\dagger_f}^2) } = {\cal C}_f(s) \; e^{i \alpha_f(s) {a^\dagger_f}^2} \; e^{i \beta_f(s) a^\dagger_f  a_f} \; e^{i \gamma_f(s) a^2_f} \;. \label{des1}
\end{equation}
Differentiating both sides of \eqref{des1} with respect to $s$ and making use of relations
\begin{equation} 
e^{i \alpha_f {a^\dagger_f}^2} \;a_f \;e^{-i \alpha_f {a^\dagger_f}^2} = a_f -2i \alpha_f ||f||^2 a^\dagger_f \;, \label{alpha}
\end{equation}
and
\begin{equation} 
e^{i \beta_f a^\dagger_f  a_f} \; a_f \;e^{-i \beta_f a^\dagger_f  a_f} = e^{-i\beta_f ||f||^2}\; a_f \;, \label{beta}
\end{equation}
we obtain the following set of differential equations 
\begin{eqnarray} 
\dot{\alpha}_f(s) & = & 1+2i ||f||^2 \alpha_f(s) {\dot \beta}_f(s) + 4 \alpha_f^2(s) {\dot \gamma}_f(s) e^{-2i \beta_f(s) ||f||^2} \;, \nonumber \\
{\dot \gamma}_f(s) & = & e^{2i \beta_f(s) ||f||^2}, \; \nonumber \\
i {\dot \beta}_f(s) & = & 4 ||f||^2 \alpha_f(s) {\dot \gamma}_f(s)  e^{-2i \beta_f(s) ||f||^2}\;, \nonumber \\
{\dot {\cal C}}_f(s) & = & 2i ||f||^4 {\cal C}_f(s) \dot{\alpha}_f(s),  \; \label{diffeq}
\end{eqnarray}
which have to be solved with the initial conditions 
\begin{equation} 
\alpha_f(0) = 0 \;, \qquad \beta_f(0) = 0 \;, \qquad \gamma_f(0) =0 \;, \qquad {\cal C}_f(0) = 1 \;. \label{cds}
\end{equation}
Upon performing the necessary algebra, we get
\begin{eqnarray} 
\alpha_f & = & \alpha_f(1) = \frac{1}{2 ||f||^2} \tanh(2 ||f||^2) \;, \nonumber \\
\beta_f & = & \beta_f(1) = \frac{i}{||f||^2} \log(\cosh(2||f||^2)) \;, \nonumber \\
{\cal C}_f & = & {\cal C}_f(1) = \left(\cosh(2 ||f||^2)\right)^{-1/2} \;, \nonumber \\
\gamma_f & =& \gamma_f(1) = \frac{1}{8 ||f||^2} \left( -4 ||f ||^2 -\sinh(4 ||f||^2) \right) \;. \label{sold}
\end{eqnarray}
As a result, the single-mode squeezed state $|\Psi \rangle$ takes the form
\begin{equation} 
|\Psi\rangle =  {\cal C}_f \sum_{n=0}^{\infty} (i \alpha_f)^n |n_f n_f\rangle \;, \qquad |n_fn_f\rangle = \frac{(a^\dagger_f)^n}{\sqrt{n!}}\frac{(a^\dagger_f)^n}{\sqrt{n!}} |\Omega_f\rangle \;. \label{sqsq}
\end{equation} 
An additional important property of the squeezed state \eqref{sqsq} (see Appendix \eqref{appB}) is derived from
\begin{equation} 
{\cal S}_f  a^{\dagger}_g {\cal S}_f^{\dagger} = a^\dagger_g - \frac{\langle f | g\rangle}{||f||^2} a^\dagger_f +  \frac{\langle f | g\rangle}{||f||^2}\left( \cosh(||f||^2) a^\dagger_f + i \sinh(||f||^2) a_f \right) \;, \label{autm1}
\end{equation} 
allowing us to demonstrate that
\begin{equation} 
{\cal S}_f  W_g {\cal S}_f^{\dagger} = W_{\bar g} \;, \label{autm2} 
\end{equation}
where $g$ denotes a generic test function supported in ${\cal W}_R$ and 
\begin{equation} 
{\bar g} =  g + \eta f \;, \label{autmm}
\end{equation}
with
\begin{eqnarray} 
\eta & = & - \frac{H(f,g)}{||f||^2} \left(1-\cosh(||f||^2) \right) + i \frac{H(f,g)}{||f||^2} \sinh(||f||^2)  \nonumber \\
& + & \frac{i}{2} \frac{\Delta_{PJ}(f,g)}{||f||^2} \left(1-\cosh(||f||^2)\right) -  \frac{1}{2} \frac{\Delta_{PJ}(f,g)}{||f||^2} \sinh(||f||^2)  \;. \label{autm3}
\end{eqnarray}
Equations~\eqref{autmm} and \eqref{autm3} indicate that the single-mode squeezing operator ${\cal S}_f$ induces an automorphism of the von Neumann algebra ${\cal M}$ \cite{Casini:2019qst}. As a result, the single-mode squeezed state $|\Psi\rangle$ is both cyclic and separating \cite{Casini:2019qst}, a crucial property for computing the relative entropy between $|\Psi\rangle$ and the vacuum state $|\Omega\rangle$. This computation will be the focus of the next section.

\section{ The Araki-Uhlmann relative entropy and the Tomita-Takesaki modular theory} \label{Autt}

To introduce the Araki-Uhlmann relative entropy, we first recall some fundamental aspects of Tomita-Takesaki modular theory \cite{Takesaki:1970aki,Witten:2018zxz,Ciolli:2019mjo,Casini:2019qst,Frob:2024ijk}.

As shown in the previous section, the states $|\Psi\rangle$ and $|\Omega\rangle$ are both cyclic and separating for the von Neumann algebra ${\cal M}$, Eq.~\eqref{vnM}. The relative anti-linear Tomita-Takesaki operator $s_{\Psi|\Omega}$ is defined as the closure of the map
\begin{equation} 
s_{\Psi |\Omega} \; a \ket{\Psi} = a^{\dagger} \; \ket{\Omega} \;, \qquad \forall a \in {\cal M} \;. \label{Stt}
\end{equation} 
The modular relative operator $ \Delta_{\Psi| \Omega}$ is obtained through the polar decomposition of $s_{\Psi| \Omega}$, namely, 
\begin{equation} 
s_{\Psi| \Omega} = J_{\Psi |\Omega} \;  \Delta_{\Psi| \Omega}^{1/2} \;, \label{pd}
\end{equation}
with $J_{\Psi|\Omega}$ being the anti-unitary relative modular conjugation. The operator $ \Delta_{\Psi |\Omega}$ is self-adjoint and positive definite.

The Araki-Uhlmann relative entropy is given by the expression 
\begin{equation} 
S(\Psi |\Omega) = - \bra{\Psi}\log \Delta_{\Psi| \Omega}  \ket{\Psi} \;.\label{ent}
\end{equation} 
As shown in \cite{Ciolli:2019mjo}, a practical approach to handling expression \eqref{ent} involves the spectral decomposition of $ \Delta_{\Psi| \Omega}$, which allows us to express it as
\begin{equation} 
\bra{\Psi}\log \Delta_{\Psi| \Omega}  \ket{\Psi} = -i \frac{d}{ds} \bra{\Psi} \Delta_{\Psi| \Omega}^{is}  \ket{\Psi} \Big|_{s=0}.
\end{equation}
Here, the unitary operator 
\begin{equation} 
\Delta_{\Psi |\Omega}^{is} = e^{i s \log \Delta_{\Psi| \Omega} } \;, \qquad s \in {\mathbb R}  \;, \label{mf}
\end{equation}
is known as the modular flow \cite{Takesaki:1970aki,Witten:2018zxz}. Accordingly, the expression for $S(\Psi |\Omega)$ is written as
\begin{equation} 
S(\Psi |\Omega) = -\frac{1}{2i} \frac{d}{ds} \left(    \bra{\Psi} \Delta_{\Psi| \Omega}^{is} -   (\Delta_{\Psi| \Omega}^{is})^\dagger \ket{\Psi} \right) \Big|_{s=0}  \;. \label{au1}
\end{equation}
A particularly useful aspect of analyzing the Araki-Uhlmann relative entropy between a single-mode squeezed state and the vacuum state lies in the remarkable relation  \cite{Casini:2019qst,Frob:2024ijk}
\begin{equation} 
\Delta_{\Psi| \Omega}^{is} = \Delta_{\Omega}^{is} \;, \label{mde}
\end{equation} 
where $\Delta_{\Omega}^{is}$ is the  flow of the Tomita-Takesaki modular operator for the vacuum state $\ket{\Omega}$ \cite{Takesaki:1970aki,Witten:2018zxz}, such that
\begin{eqnarray} 
s_{\Omega} \; a |\Omega\rangle &  =  & a^{\dagger} |\Omega\rangle \;, \qquad \forall a \in {\cal M}  \;, \nonumber \\
s_{\Omega} & = & J_{\Omega} \Delta_{\Omega}^{1/2}  \;. \label{som}
\end{eqnarray}
The validity of Eq.~\eqref{mde} stems from the unitarity of the single-mode squeezing operator ${\cal S}_f$. To see this explicitly, consider
\begin{equation}
s_{\Psi| \Omega} \;a |\Omega\rangle = s_{\Psi| \Omega} (a {\cal S}_f^\dagger) {\cal S}_f |\Omega\rangle = s_{\Psi| \Omega}(a {\cal S}_f^\dagger) |\Psi\rangle = {\cal S}_f\; a^\dagger |\Omega\rangle = {\cal S}_f s_{\Omega} \;a |\Omega\rangle \;, \label{feq}
\end{equation}
which leads to $\left( s_{\Psi| \Omega} - {\cal S}_f s_{\Omega} \right) \;a |\Omega\rangle =0 \;$ for any element $a \in {\cal M}$, {\it i.e.}, $s_{\Psi| \Omega} =  {\cal S}_f \;s_{\Omega}$.
Moreover, 
\begin{equation} 
\Delta_{\Psi |\Omega} = s_{\Psi| \Omega}^\dagger \;s_{\Psi| \Omega} = s_{\Omega}^\dagger {\cal S}_f^\dagger {\cal S}_f \;s_{\Omega}= \Delta_{\Omega} \;. 
\end{equation} 
We now proceed to evaluate the relative entropy.

\section{Evaluation of the Araki-Uhlmann relative entropy}\label{comp}

As we will see, evaluating the relative entropy $S(\Psi |\Omega)$ involves analyzing the two-point function 
$\langle \Omega |\; {\cal S}_f {\cal S}_g \; |\Omega\rangle$, where $f$ and $g$ are smooth test functions supported in ${\cal W}_R$. The explicit expression of the two-point function reads 
\begin{eqnarray} 
\langle \Omega |\; {\cal S}_f {\cal S}_g \; |\Omega\rangle & = & {\cal C}_f {\cal C}_g\; \langle \Omega |\; e^{i \alpha_f (a^\dagger_f)^2} e^{i \beta_f a^\dagger_f a_f} e^{i \gamma_f a^2_f} \;e^{i \alpha_g (a^\dagger_g)^2} e^{i \beta_g a^\dagger_g a_g} e^{i \gamma_g a^2_g} \;|\Omega\rangle \nonumber \\
& =&  {\cal C}_f {\cal C}_g \;\langle \Omega |\;e^{i \gamma_f a^2_f} \;e^{i \alpha_g (a^\dagger_g)^2}\;|\Omega\rangle \nonumber\\
&=&{\cal C}_f {\cal C}_g \;
\sum_{n=0}^{\infty} \frac{(-\gamma_f \alpha_g)^n}{n!  n!} \langle\Omega|\;  \left[ (a^2_f)^n, ({a^\dagger_g}^2)^n\right] \; |\Omega \rangle \;. \label{sm}
\end{eqnarray}
Employing the relation
\begin{equation} 
\left[ (a^2_f)^n, ({a^\dagger_g}^2)^n\right] = \sum_{k=1}^{n} k! {\binom{n}{k}}^2 \langle f |g\rangle^n (a^\dagger_g)^{n-k} (a_f)^{n-k} \;, \label{bin}
\end{equation}
we find
\begin{equation} 
\langle \Omega|\; \left[ (a^2_f)^n, ({a^\dagger_g}^2)^n\right] \;|\Omega \rangle = (2n)! \;\langle f|g\rangle^{2n} \;, \label{ff2}
\end{equation}
so that 
\begin{equation} 
\langle \Omega |\; {\cal S}_f {\cal S}_g \; |\Omega\rangle = {\cal C}_f {\cal C}_g \sum_{n=0}^{\infty}  \frac{(-\gamma_f \alpha_g)^n}{n!  n!} (2n)! \;\langle f|g\rangle^{2n} \;. \label{ff3}
\end{equation}
This expression can be evaluated to obtain the closed-form result:
\begin{equation} 
\langle \Omega |\; {\cal S}_f {\cal S}_g \; |\Omega\rangle = {\cal C}_f {\cal C}_g \frac{1}{\sqrt{ 1 + 4 \alpha_g \gamma_f \langle f|g\rangle^2}} \;. \label{ffsum}
\end{equation} 
Let us now return to the relative entropy. Recalling that $|\Psi\rangle = {\cal S}_f |\Omega\rangle$, and making use of \cite{Casini:2019qst,Frob:2024ijk}
\begin{equation} 
\Delta_{\Omega}^{is}\; {\cal S}_f \; \Delta_{\Omega}^{-is} = {\cal S}_{f_s} \;, \label{da}
\end{equation}
where, from the Bisognano-Wichmann results \cite{Bisognano:1975ih},  
\begin{equation} 
f_s(x)= f(\Lambda_{-s} x) \;, \label{fst}
\end{equation}
with $\Lambda_s$ denoting a Lorentz boost: 
\begin{align}
\Lambda_s: 
\left\{
    \begin {aligned}
         &  \;\; x'  =  \cosh(2\pi s) \;x - \sinh(2 \pi s) \; t, \\
         &  \;\;  t'  =  \cosh(2\pi s) \;t - \sinh(2 \pi s) \; x,             
    \end{aligned}
\right. \label{bst}
\end{align}
it follows that 
\begin{equation} 
\langle \Psi|\; \Delta_{\Psi \Omega}^{is} \;|\Psi\rangle  =  \langle \Psi|\; \Delta_{ \Omega}^{is} \;|\Psi\rangle = \langle \Omega |\; {\cal S}_f^{\dagger} \Delta_{ \Omega}^{is} {\cal S}_f \;|\Omega\rangle = \langle \Omega |\; {\cal S}_f^{\dagger} \Delta_{ \Omega}^{is} {\cal S}_f  \Delta_{\Omega}^{-is} \;|\Omega\rangle = \langle \Omega|\; {\cal S}_f^{\dagger} {\cal S}_{f_{s}}\;|\Omega \rangle \;. \label{qr1}
\end{equation} 
In the last expression, we used the fact that the vacuum is invariant under the action of $\Delta_{\Omega}^{is}$, \ie,\ $\Delta_{ \Omega}^{is}\; |\Omega \rangle = |\Omega \rangle.$
Consequently, the Araki-Uhlmann relative entropy satisfies the final relation 
\begin{equation} 
S(\Psi |\Omega) = -\frac{1}{2i} \frac{d}{ds} \left(    \bra{\Psi}( \Delta_{\Psi| \Omega}^{is} -  \textrm{c.c.} )\ket{\Psi} \right) \Big|_{s=0} = - \frac{\alpha_f^2 {\cal C}_f^2}{(1- 4 \alpha_f^2||f||^4)^{3/2}} ||f||^2 \Delta_{PJ}\left(f,f'_s\Big|_{s=0}\right)\;. \label{qrf}
\end{equation}
with 
\begin{equation} 
{\cal C}_f = \left( \cosh(2 ||f||^2) \right)^{-1/2} \;, \qquad  \alpha_f = \frac{1}{2||f||^2} \tanh(2||f||^2)  \;, \label{auxf}
\end{equation}
and 
\begin{equation} 
f'_s\Big|_{s=0} = \frac{d}{ds} f_s \Big|_{s=0} \;. \label{dfs}
\end{equation}
Expression \eqref{qrf} represents the central result of this work, providing a closed-form analytic expression for the relative entropy between the single-mode squeezed
state $|\Psi\rangle$ and the vacuum state $|\Omega\rangle$.

At this stage, a few observations can be made: As in the case of coherent states \cite{Ciolli:2019mjo,Casini:2019qst,Frob:2024ijk,Guimaraes:2025cqt}, the Araki-Uhlmann relative entropy between a single-mode squeezed state  $|\Psi\rangle$ and the vacuum state $|\Omega\rangle$ is entirely determined by the smeared Pauli-Jordan distribution. Moreover, it is worth noting that expression \eqref{qrf} can be
written as
\begin{equation}
    S(\Psi |\Omega)  =  \frac{2\,\alpha_f^2 {\cal C}_f^2}{(1- 4 \alpha_f^2||f||^4)^{3/2}} ||f||^2 S_{\rm coherent}\;,
\end{equation}
where $S_{\rm coherent}$ stands for the relative entropy between
the coherent state $\e^{i\varphi(f)}\ket{0}$ and the vacuum state, see Eq.~(19) of \cite{Guimaraes:2025cqt}. As such,
one can make use of the analysis done in \cite{Casini:2019qst} and express $S(\Psi|\Omega)$ as an integral of the smeared energy-momentum tensor, see Appendix A.3 of \cite{Casini:2019qst}. This property leads to the expected behavior of the relative entropy \cite{Ciolli:2019mjo,Casini:2019qst,Frob:2024ijk,Guimaraes:2025cqt}: {\it i)} it is positive, {\it ii)} it decreases as the mass parameter increases, and {\it iii)} it increases with the size of the space-time region considered.

\section{Conclusion}\label{conc}

In this work, we have derived a closed-form expression for the Araki-Uhlmann relative entropy between a single-mode squeezed state and the vacuum state in a free massive scalar QFT. Utilizing the framework of Tomita-Takesaki modular theory and the Bisognano-Wichmann theorem, we demonstrated that the modular flow associated with the single-mode squeezed state coincides with that of the vacuum, a consequence of the unitarity of the single-mode squeezing operator. This key observation enabled us to compute the relative entropy analytically, revealing that it is governed entirely by the smeared Pauli-Jordan distribution.

Our results extend previous findings on coherent states and confirm that the relative entropy between a single-mode squeezed state and the vacuum state preserves its expected structural features: it is strictly positive, increases with the size of the space-time region under consideration, and decreases as the mass parameter grows. These behaviors underscore the utility of relative entropy
as a robust measure of quantum distinguishability and information content for single-mode squeezed states in relativistic quantum field-theoretic settings.

The techniques and results presented here enhance the understanding of quantum information measures in QFT, particularly for single-mode squeezed states, and provide a concrete example where the intricate algebraic structures of QFT yield exact, physically meaningful quantities. Future investigations
may extend these methods to more general classes of states, such as multi-mode squeezed states or states in interacting theories, further elucidating the interplay between algebraic modular theory and quantum information in relativistic systems.

\section*{Acknowledgments}
The authors would like to thank the Brazilian agencies Conselho Nacional de Desenvolvimento Científico e Tecnológico (CNPq), Coordenação de
Aperfeiçoamento de Pessoal de Nível Superior - Brasil (CAPES) and Fundação Carlos Chagas Filho de Amparo à Pesquisa do Estado do Rio de Janeiro (FAPERJ) for financial support. In particular, A. F.~Vieira is supported by a postdoctoral grant from FAPERJ in the Pós-doutorado Nota 10 program, grant No. E-
26/200.135/2025. S. P.~Sorella, I.~Roditi, and M. S.~Guimaraes are CNPq researchers under contracts 301030/2019-7, 311876/2021-8, and 309793/2023-8, respectively. 

%\end{acknowledgments}

\appendix

\section{The massive real scalar field in $1+1$ Minkowski spacetime}\label{appA}
In this appendix, we summarize the key aspects of the canonical quantization of a massive real scalar field in $1+1$-dimensional Minkowski spacetime.
\subsection{ Field expansion and commutation relations}

The massive scalar field $\varphi(t,x)$ can be expressed in terms of plane waves as:
\begin{equation} \label{qf}
\varphi(t,x) = \int \! \frac{d k}{2 \pi} \frac{1}{2 \omega_k} \left( e^{-ik_\mu x^\mu} a_k + e^{ik_\mu x^\mu} a^{\dagger}_k \right), 
\end{equation} 
where $\omega_k  = k^0 = \sqrt{k^2 + m^2}$ is the relativistic energy dispersion relation. The field operators satisfy the canonical commutation relations:
\begin{align}
[a_k, a^{\dagger}_q] &= 2\pi \, 2\omega_k \, \delta(k - q), \\ \nonumber 
[a_k, a_q] &= [a^{\dagger}_k, a^{\dagger}_q] = 0. 
\end{align}
Since quantum fields are operator-valued distributions \cite{Haag:1992hx}, they must be smeared with test functions to produce well-defined operators in Hilbert space. This is done by defining the smeared field operator as:
\begin{align} 
\varphi(h) = \int \! d^2x \; \varphi(x) h(x)=a_h+a_h^\dagger \;, \label{smmd}
\end{align}
where $h(x)$ is a real smooth test function with compact support and
\begin{align}
    a_h&=\int\frac{\dd k}{2\pi}\frac{1}{2\omega_k}h(\omega_k,k)a_k,\\
    a_h^\dagger&=\int\frac{\dd k}{2\pi}\frac{1}{2\omega_k}h^*(\omega_k,k)a_k^\dagger.
\end{align}

\subsection{Inner product and two-point functions}

With the smeared fields, the Lorentz-invariant inner product between two test functions $f(x)$ and $g(x)$ in the vacuum state is introduced by means of the two-point smeared Wightman function
\begin{align} \label{InnerProduct}
\langle f \vert g \rangle &= \langle 0 \vert \varphi(f) \varphi(g) \vert 0 \rangle =  \frac{i}{2} \Delta_{PJ}(f,g) +  H(f,g) \;, 
\end{align}
where $ \Delta_{PJ}(f,g)$ and $H(f,g)$ are the smeared versions of the Pauli-Jordan and Hadamard distributions, respectively. These are defined as:
\begin{align}
\Delta_{PJ}(f,g) &=  \int \! d^2x d^2y f(x) \Delta_{PJ}(x-y) g(y) \;,  \nonumber \\
H(f,g) &=  \int \! d^2x d^2y f(x) H(x-y) g(y)\;. \label{mint}
\end{align}
The Pauli-Jordan function $\Delta_{PJ}(x-y)$ and the Hadamard function $H(x-y)$ take the explicit forms:
\begin{eqnarray} 
\Delta_{PJ}(t,x) & =&  -\frac{1}{2}\;{\rm sign}(t) \; \theta \left( \lambda(t,x) \right) \;J_0 \left(m\sqrt{\lambda(t,x)}\right) \;, \nonumber \\
H(t,x) & = & -\frac{1}{2}\; \theta \left(\lambda(t,x) \right )\; Y_0 \left(m\sqrt{\lambda(t,x)}\right)+ \frac{1}{\pi}\;  \theta \left(-\lambda(t,x) \right)\; K_0\left(m\sqrt{-\lambda(t,x)}\right) \;, \label{PJH}
\end{eqnarray}
where 
\begin{equation} 
\lambda(t,x) = t^2-x^2 \;, \label{ltx}
\end{equation}
and $(J_0,Y_0,K_0)$ are Bessel functions, while $m$ is the mass parameter.

\subsection{Physical interpretation and causality}

Both the Hadamard and Pauli–Jordan distributions are Lorentz-invariant, but they fulfill distinct roles in the structure of quantum field theory. The Pauli–Jordan distribution, $\Delta_{PJ}(x)$, is particularly significant in encoding relativistic causality, as it vanishes for spacelike-separated arguments, thereby ensuring the commutativity of field operators at spacelike separation. Moreover, $\Delta_{PJ}(x)$ is antisymmetric under spacetime inversion, $x \mapsto -x$, in contrast to the Hadamard distribution $H(x)$, which is symmetric.

In the smeared-field formalism, the commutator of scalar field operators is given by
\[
\left[\varphi(f), \varphi(g)\right] = i \Delta_{PJ}(f, g).
\] 
where $f(x)$ and $g(x)$ are test functions. Consequently, if the supports of $f(x)$ and $(g(x)$ are spacelike-separated, then
\[
\left[\phi(f), \phi(g)\right] = 0,
\]
reflecting the microcausality condition in a manifestly covariant framework.

\section{Some details about the single-mode squeezed states in QFT} \label{appB}

In this appendix, we aim to proof that the single-mode squeezing operator $\mathcal{S}_f$ defines an automorphism for the von Neumann algebra $\M$. We start with the single-mode squeezing operator\footnote{We have included a factor of 1/2 for convenience, but it does not spoil our proof.}
\begin{equation}
    \mathcal{S}_f=e^{\mathcal{A}_f}, \quad \mathcal{A}_f=\frac{i}{2}(a_f^2+{a_f^{\dagger}}^2),
\end{equation}
and the Weyl operator
\begin{equation}
    W_g=e^{i(a_g+ a_g^\dagger)}=e^{-\frac{1}{2}||g||^2} e^{ia_g^\dagger}\,e^{ia_g},
\end{equation}
where $(f,g)$ are smooth test functions with support in the right Rindler wedge and $||g||^2= \bra{g}\ket{g}.$ Then, it follows that
\begin{equation}
    \S_f \,W_g \,\S_f^\dagger=e^{-\frac{1}{2}||g||^2}(\S_f \,e^{i a_g^\dagger}\,\S_f^\dagger)(\S_f \,e^{i a_g}\,\S_f^\dagger).
\end{equation}
The term $ \S_f \,e^{i a_g^\dagger}\,S_f^\dagger$ can be expanded as
\begin{align}
    \S_f \,e^{i a_g^\dagger}\,S_f^\dagger&=\sum_{n=0}^{\infty}\frac{i^n}{n!}\S_f (a_g^\dagger)^n \S_f^\dagger,
\end{align}
with
\begin{align}
\S_f\,a_g^\dagger\,S_f^\dagger&=e^{\A}\,a_g^\dagger\, e^{-\A}\nonumber\\
    &=\sum_{n=0}^{\infty}\frac{1}{n}[\A,[\A,\dots[\A,g^\dagger_g]\,]\,].
\end{align}
We can use the relations
\begin{align}
    [\A,a_g^\dagger]&=i \bra{f}\ket{g}a_f,\\
    [\A,[\A,a_g^\dagger]\,]&=\bra{f}\ket{f}||f||^2 a_f^\dagger,\\
    [\A,[\A,[\A,a_g^\dagger]\,]\,]&=i\bra{f}\ket{g}||f||^4 a_f,\\
    [\A,[\A,[\A,[\A,a_g^\dagger]\,]\,]\,]&=\bra{f}\ket{g}||f||^6 a_f^\dagger,
\end{align}
    to recast $ \S_f \,e^{i a_g^\dagger}\,S_f^\dagger$ into a more manageable form. Explicitly, it reads
    \begin{align}
        \S_f \,a_g^\dagger\,S_f^\dagger&=a^\dagger_g-\frac{\bra{f}\ket{g}}{||f||^2}a^\dagger_f+\frac{\bra{f}\ket{g}}{||f||^2}\left(\cosh{||f||^2}a_f^\dagger+i \sinh{||f||^2}a_f  \right)\nonumber\\
        &=\sigma^\dagger.
    \end{align}
It is straightforward to see that $\S_f(a_g^\dagger)^n \S_f^\dagger=(\sigma^\dagger)^n$ and $\S_f(a_g)^n \S_f^\dagger=(\sigma)^n$. These relations allow us to write
\begin{align}
    \S_f\,e^{i a_g^\dagger}\,\S_f^\dagger&=\sum_{n=0}^{\infty}\frac{i^n}{n!}(\sigma^\dagger)^n=e^{i\sigma^\dagger},\\
    \S_f\,e^{i a_g}\,\S_f^\dagger&=\sum_{n=0}^{\infty}\frac{i^n}{n!}\sigma^n=e^{i\sigma},
\end{align}
both of which ultimately lead us to
\begin{equation}\label{relation1}
    \S_f\,W_g\, S_f^\dagger=e^{-\frac{1}{2}||g||^2}e^{i\sigma^\dagger}e^{i\sigma}.
\end{equation}
Since the $\sigma$ operator is linear in the creation and annihilation operators, Eq. \eqref{relation1} can be rewritten with the help of the Baker-Campbell-Hausdorff formula as
\begin{equation}
    \S_f\,W_g\, S_f^\dagger=e^{-\frac{1}{2}||g||^2}e^{i(\sigma+\sigma^\dagger)-\frac{1}{2}\comm{\sigma^\dagger}{\sigma}}.
\end{equation}
We note that the commutator term is a $c$-number which can be easily calculated to produce $\comm{\sigma^\dagger}{\sigma}=-||g||^2$, where the inner product \eqref{InnerProduct} and its complex conjugate have been used. Lastly, by using Eq.~\eqref{smmd}, the sum $\sigma+\sigma^\dagger$ is given by
\begin{equation}
    \sigma+\sigma^\dagger=\phi(h)=\phi(g)+\gamma a_f+ \gamma^* a_f^\dagger
\end{equation}
where
\begin{align}
    \gamma&=-\frac{H(f,g)}{||f||^2}(1-\cosh{||f||^2})+\frac{i}{2}\frac{\Delta(f,g)}{||f||^2}(1-\cosh{||f||^2})\nonumber\\
    &+i\frac{H(f,g)}{||f||^2}\sinh{||f||^2}-\frac{1}{2}\frac{\Delta(f,g)}{||f||^2}\sinh{||f||^2}.
\end{align}
Putting all the pieces together, we arrive at our final result, which reads
\begin{align}
    \S_f\,W_g\,\S_f^\dagger&=\eta W_h,\\
    \eta&=e^{-\frac{1}{2}||g||^2}e^{-\frac{1}{2}\comm{\sigma^\dagger}{\sigma}}=1,
\end{align}
with $h= g+\gamma f.$ This completes the proof that the single-mode squeezing operator $\S_f$ defines an automorphism for the von Neumann algebra $\mathcal{M}$.
\bibliography{refs}

\end{document}